# Cross-stream migration characteristics of a deformable droplet in a non-isothermal Poiseuille Flow through Microfluidic Channel

Sayan Das, Shubhadeep Mandal and Suman Chakraborty†

Department of Mechanical Engineering, Indian Institute of Technology Kharagpur,
Kharagpur – 721302, India

The migration characteristics of a suspended deformable droplet in a parallel plate microchannel is studied, both analytically and numerically, under the combined influence of a constant temperature gradient in the transverse direction and an imposed pressure driven flow. Any predefined transverse position in the micro channel can be attained by the droplet depending on the applied temperature gradient in the cross-stream direction or how small the droplet is with respect to the channel width. For the analytical solution, an asymptotic approach is used, where we neglect any effect of inertia or thermal convection of the fluid in either of the phases. To obtain a numerical solution, we use the conservative level set method. Variation of temperature in the flow field causes a jump in the tangential component of stress at the droplet interface. This jump in stress component, which is the thermal Marangoni stress, is an important factor that controls the trajectory of the droplet. The direction of cross-stream migration of the droplet is decided by the magnitude of the critical Marangoni stress, corresponding to which the droplet remains stationary. In order to analyze practical microfluidic setup, we do numerical simulations where we consider wall effects as well as the effect of thermal convection and finite shape deformation on the cross-stream migration of the droplet. Increase in the thermal Marangoni stress shifts the steady state transverse position of the droplet further away from the channel centerline, for any particular value of capillary number (which signifies the ratio of the viscous force to the surface tension force). The confinement ratio, which is the ratio of the droplet radius to the channel height, plays an important role in predicting the transverse position of droplet. A large confinement ratio drives the droplet towards the channel center whereas smaller confinement ratios cause the droplet to move towards the wall. The thermal conductivity ratio (ratio of the thermal conductivity of the droplet phase to that of the carrier phase) is also found to be an important factor towards manipulating the lateral position of the droplet.

# 1. Introduction

Droplet dynamics has been a progressively rising topic of interest due to its wide gamut of applications in various microfluidic devices (Baroud et al. 2010; Seemann et al. 2012; Stone et al. 2004). Some of these applications are: cell encapsulation, drug delivery, analytical detection and reagent mixing (Teh et al. 2008; Huebner et al. 2008; Di Carlo et al. 2007; Zhu & Fang 2013; Baroud et al. 2010). The transport of droplets and bubbles also finds it applications in manufacturing of pure materials (Murr & Johnson 2017; Kim et al. 2014; Anna 2016; Sajeesh & Sen 2014). With the advent of space exploration, the studies on droplet dynamics under micro gravity environment is becoming more and more important (Das et al. 2017; Pak et al. 2014; Balasubramaniam & Lavery 1989; Subramanian 1983). The motion of the droplets in this scenario is solely due to variation in interfacial tension or due to any imposed flow, if present. Also, the lateral migration of droplets as well as their steady state position in a channel gives an idea of the distribution of droplets, which in turn is of great importance in determining the flow rate though the channel for a given pressure drop (Kinoshita et al. 2007; Ward et al. 2005; Anna 2016).

It has been shown that an eccentrically placed droplet in a Poiseuille flow field undergoes migration only in the axial direction in the absence of droplet deformation, fluid inertia, surfactants, fluid viscoelasticity and variation in temperature of the flow field (Hetsroni & Haber 1970). In the Stokes flow limit, various theoretical studies have shown that the presence of any of the above mentioned effects causes the cross-stream migration of droplets in a pressure driven flow (Stan et al. 2011; Bandopadhyay et al. 2016; Mandal et al. 2015; Pak et al. 2014; Chan & Leal 1979; Wu & Hu 2011; Nas & G. Tryggvason 2003). Haber and Hetsroni analytically studied the cross-stream migration of a spherical droplet suspended in an arbitrary flow field (Hetsroni & Haber 1970). Chan and Leal (1979) obtained closed form solutions for the cross-stream migration velocity of a nearly spherical droplet suspended in another fluid undergoing Poiseuille flow. The diameter of the droplet was assumed to be much smaller than the channel width to neglect any wall effects. It was shown that the droplet migrated towards the channel centerline for $0.5 < \lambda < 10$, whereas it migrated away from the centerline for other values of $\lambda$, where $\lambda$ is the ratio of the droplet viscosity to the viscosity of the suspending phase. The axial velocity of the droplet as predicted by the theory always lagged behind the undisturbed Poiseuille velocity field. Other than these, two-dimensional numerical simulations using the boundary integral method have also been done to study the migration of deformable droplets in a pressure driven flow at zero Reynolds number (Zhou & Pozrikidis 1994; Zhou & Pozrikidis 1993). They showed that for drops with viscosity ratio 10, which are initially close to the centerline migrates towards the wall and those close to wall move towards the channel centerline. Mortazavi and Tryggvason (2000) studied numerically the cross-stream migration of a deformable droplet in a two-dimensional Poiseuille flow (Mortazavi & Tryggvason 2000). They did their analysis for the limiting case of low as well as high Reynolds number. Under low Reynolds number limit, they showed that for $\lambda = 0.125$, the droplet always migrates towards the centerline of the channel, whereas for $\lambda = 1$, the droplet moved away from the centerline of the channel. They showed that the migration rate increased when deformation of the droplet was taken into consideration.

The use of external effects like electric, (Ahn et al. 2006; Link et al. 2006; Mandal et al. 2016) acoustic (Seemann et al. 2012), thermal (Karbalaei et al. 2016; Young et al. 1959; Balasubramaniam & Subramaniam 1996) and magnetic effects (Seemann et al. 2012) has been of great importance in recent times due to its wide application in variety of fields. The presence of these effects introduces a jump in stress at the droplet interface which alters the net hydrodynamic force acting on the droplet and hence the droplet dynamics changes. In the present study, we have focused on the effect of temperature variation in the flow field. The variation of temperature along the droplet surface develops a Marangoni stress, which drives the droplet even in a quiescent medium. This is called thermocapillary migration and was first demonstrated experimentally and theoretically by Young et al.(1959). Followed by the study done by Young et al., the effect of several aspects like fluid inertia, (Haj-Hariri et al. 1990) droplet deformation, (Nadim et al. 1990) bounding wall (Meyyappan & Subramanian 1987; Barton & Shankar Subramanian 1990; Barton & Subramanian 1991; Chen 1999; Chen 2003) and thermal convection (Balasubramaniam & Subramanian 2004; Yariv & Shusser 2006; Zhang et al. 2001) on the thermocapillary migration of a droplet suspended in a quiescent medium was studied. Wang et al. and Hariri et al. preformed three dimensional numerical simulations to study the thermocapillary migration of deformable as well as non-deformable droplets for finite values of thermal Marangoni number $(Ma_T)$, which signifies the ratio of the thermally induced Marangoni stress to the viscous stresses acting on the droplet (Wang et al. 2008; Haj-Hariri et al. 1997). Balasubramanian and Subramanian (1996) found out that for large values of $Ma_T$, the droplet velocity increases with increase in $Ma_T$. In a recent work, Raja Shekhar and co-workers, have theoretically shown the combined effect of imposed Poiseuille flow as well as the thermal Marangoni stress on the droplet dynamics (Sharanya & Raja Sekhar 2015; Choudhuri & Raja Sekhar 2013). They in their work have showed that in the absence of surfactants, fluid inertia and droplet deformation, a linear combination of the migration velocity due to thermocapillary effect with that due to the imposed flow gives us the final droplet migration velocity.

A lot of work has been done to show the effect of temperature variation as well as imposed flow on the droplet migration characteristics, both theoretically and numerically. However, no work is reported in the literature that shows the effect of a transverse temperature gradient on the cross-stream migration of a droplet in a pressure driven flow inside a microchannel. In this work, we have obtained an asymptotic solution of the flow field and the temperature field assuming small deformation of the droplet. We have then performed two-dimensional numerical simulations to show the effect of different dimensionless parameters on the migration characteristics of droplet.

## 2. Problem formulation

### 2.1. Physical system

The present problem deals with a viscous, Newtonian droplet of radius $a$, having density $\rho_i$, thermal conductivity $k_i$ and viscosity $\mu_i$. This droplet is suspended in another Newtonian fluid of thermal conductivity $k_e$, density $\rho_e$ and viscosity $\mu_e$ in a parallel plate micro channel. The subscript '$e$' and '$i$' signify the carrier phase and droplet phase properties or quantities respectively.

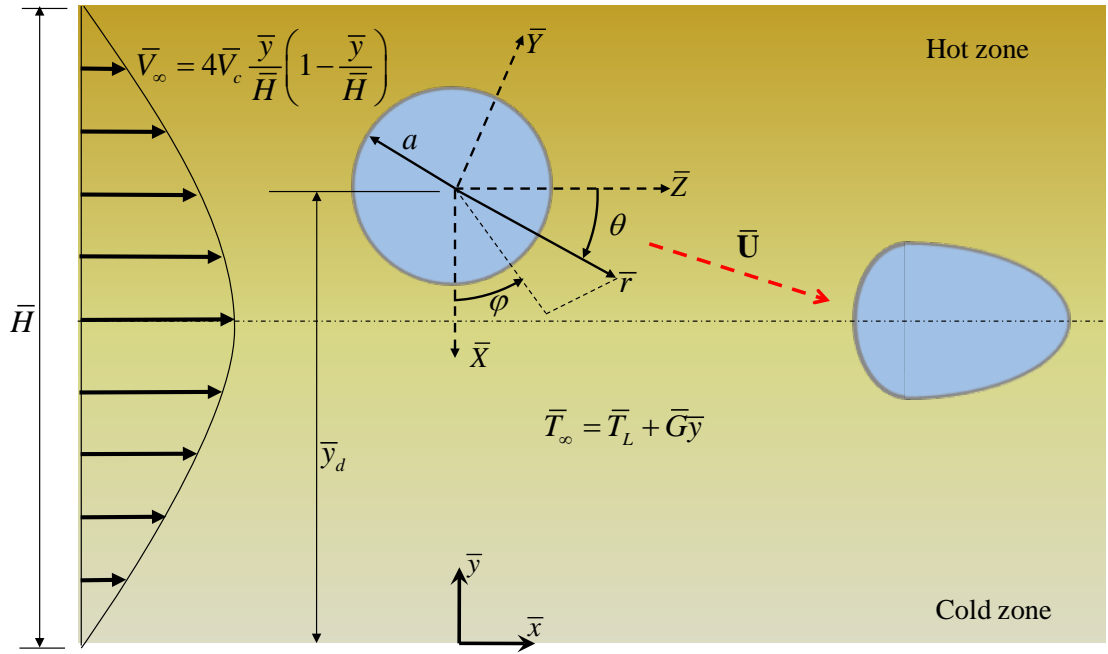

FIGURE 1. Schematic of a Newtonian droplet suspended in a plane Poiseuille flow. $\bar{H}$ is the total width of the flow field. The droplet has a radius of $a$ and is placed at a position $\bar{y}_d$ from the $\bar{x}$-axis. Both spherical coordinates $(\bar{r},\theta,\varphi)$ and cartesian coordinates $(\bar{x},\bar{y})$ are shown in the above figure. The spherical coordinate is attached to the centroid of the droplet, while the cartesian coordinate is attached to the bottom plate. A separate set of Cartesian coordinates $(\bar{X},\bar{Y},\bar{Z})$ is also attached to the centroid of the undeformed droplet.

The droplet is placed in an off-center position, above the centerline of the micro channel as can be seen from the schematic in figure 1. A rectangular Cartesian coordinate system $(\bar{x},\bar{y})$ is adopted, which is attached to the bottom plate (figure 1). For the purpose of theoretical analysis, the droplet is assumed to be much smaller as compared to the width of the micro channel. Hence the suspending medium in the schematic above is effectively shown to be unbounded. In figure 1, $\bar{y}_d$ represents lateral position of the droplet measured from the $\bar{x}$-axis. A background pressure driven flow is applied to the carrier phase in the axial direction ($\bar{x}$-direction). The temperature of the upper as well as the lower plate, though held constant, are different from each other such that there exists a linearly increasing temperature field (in the absence of droplet) in the cross-stream direction. Thus the temperature field $(\bar{T})$ can be expressed as $\bar{T} = \bar{T}_L + \bar{G}\bar{y}$, where $\bar{G}$ is the constant temperature gradient and $\bar{T}_L$ is the reference temperature or the temperature of the lower plate $(\bar{y} = 0)$. All the material properties of the two phases are assumed to be constant except the surface tension, $\bar{\sigma}$. The surface tension depends on the temperature distribution at the droplet interface $(\bar{T}_s)$. The surface tension corresponding to the reference temperature, $\bar{T}_L$, is called the equilibrium surface tension and is denoted by $\bar{\sigma}_{eq}$. It is to be noted that all dimensional variables are denoted by a 'overbar' at their top, whereas those missing a 'overbar' represent dimensionless quantities.

## 2.2. Governing equations and boundary conditions

We first state the general governing differential equations and the relevant boundary conditions for the temperature as well as the flow field. The governing equations for solving the flow field are the continuity and the Navier-Stokes equations. The continuity and the Navier-Stokes equations for the droplet phase are written below

$$\left. \begin{array}{l} \bar{\nabla} \cdot \bar{\mathbf{u}}_i = 0, \\ \rho_i \left\{ \dfrac{\partial \bar{\mathbf{u}}_i}{\partial \bar{t}} + \bar{\nabla} \cdot (\bar{\mathbf{u}}_i \bar{\mathbf{u}}_i) \right\} = -\bar{\nabla}\bar{p}_i + \mu_i \bar{\nabla} \cdot \left\{ \bar{\nabla}\bar{\mathbf{u}}_i + (\bar{\nabla}\bar{\mathbf{u}}_i)^T \right\}, \end{array} \right\} \quad (1)$$

where $\mathbf{u}_i$ and $p_i$ are the velocity and the pressure field of the droplet phase and $\bar{t}$ represents the dimensional time. The governing equations, for solving the flow field of the suspending phase, are mentioned below

$$\left. \begin{array}{l} \bar{\nabla} \cdot \bar{\mathbf{u}}_e = 0, \\ \rho_e \left\{ \dfrac{\partial \bar{\mathbf{u}}_e}{\partial \bar{t}} + \bar{\nabla} \cdot (\bar{\mathbf{u}}_e \bar{\mathbf{u}}_e) \right\} = -\bar{\nabla}\bar{p}_e + \mu_e \bar{\nabla} \cdot \left\{ \bar{\nabla}\bar{\mathbf{u}}_e + (\bar{\nabla}\bar{\mathbf{u}}_e)^T \right\}, \end{array} \right\} \quad (2)$$

where $\mathbf{u}_e$ and $p_e$ are the velocity and the pressure field of the suspending phase. The governing equation for the temperature field is the energy equation and is written for both the phases as follows

$$\left.\begin{aligned}\rho_i c_p \left( \frac{\partial \overline{T}_i}{\partial \overline{t}} + \overline{\mathbf{u}}_i \cdot \overline{\nabla} \overline{T}_i \right) &= k_i \overline{\nabla}^2 \overline{T}_i, \\ \rho_e c_p \left( \frac{\partial \overline{T}_e}{\partial \overline{t}} + \overline{\mathbf{u}}_e \cdot \overline{\nabla} \overline{T}_e \right) &= k_e \overline{\nabla}^2 \overline{T}_e, \end{aligned}\right\} \quad (3)$$

where $\overline{T}_i$, $\overline{T}_e$ represents the temperature field inside and outside the droplet respectively and $c_p$ is the specific heat capacity for either of the phases. In the above equation we have neglected the presence of any viscous dissipation in either of the phases. The boundary conditions for the flow field at the droplet surface consists of the kinematic condition, the tangential velocity continuity condition and the stress balance condition. The interfacial boundary conditions for the energy equation are the continuity of temperature and heat flux at the droplet surface. Both the temperature as well as the flow field must satisfy the far-field conditions and should also be bounded at the center of the droplet. These boundary conditions are stated in detail in section 3.

Other than these, the initial conditions required to solve the above governing differential equations are

$$\left.\begin{aligned}\text{at } \overline{t} = 0, \quad &\overline{T} = \overline{T}_L + \overline{G}\overline{y}, \\ \text{at } \overline{t} = 0, \quad &\overline{u} = 4\overline{V}_c \left( \frac{\overline{y}}{\overline{H}} \right)\left( 1 - \frac{\overline{y}}{\overline{H}} \right), \overline{v} = 0. \end{aligned}\right\} \quad (4)$$

The above governing equations, being coupled non-linear equations cannot be solved analytically. However, the same can be solved numerically with the help of a finite element based method. In order to solve the equations analytically further simplifications are needed to be done with the help of some assumptions. We thus first deal with the theoretical approach where we asymptotically solve the governing equations along with the corresponding boundary conditions. Then we move towards solving the above governing equations numerically.

## 3. Theoretical approach

We now discuss about the asymptotic analysis done to solve the governing equations for flow and temperature field. We adopt a regular perturbation methodology to fulfill our objective.

### *3.1. Assumptions*

We perform an asymptotic analysis to obtain a theoretical solution of the flow field and the thermal field. The assumptions used for this study, in order to simplify the governing equations and boundary conditions are: (i) The droplet is assumed to be neutrally buoyant $\left(\rho_i = \rho_e = \rho\right)$ and the droplet radius is assumed to be much less than the channel width $\bar{H}$. (ii) Any heat transfer in the fluid is assumed to be mainly by conduction, that is the thermal Péclet number ($Pe_T = \bar{V}_c a / \alpha_e$ where $\bar{V}_c$ is the centerline velocity of imposed Poiseuille flow and $\alpha_e$ is the thermal diffusivity of the suspending fluid) is assumed to be small. $Pe_T$ physically signifies the relative strength of heat transfer by convection with respect to that by conduction. (iii) The flow field is assumed to be driven by viscous and pressure forces. The effect of inertia is neglected, so that the Reynolds number ($Re = \rho \bar{V}_c a / \mu_e$) is small. $Re$ physically signifies the ratio of inertia to the viscous forces in fluid flow. (iv) The surface tension is assumed to be linearly dependent on the temperature distribution at the interface of the droplet. (v) We consider the surface tension force to be dominant over the viscous forces so that only small deformation of the droplet takes place. In other words we assume a low value of capillary number $\left(Ca = \mu_e \bar{V}_c / \bar{\sigma}_{eq}\right)$ which signifies the relative strength of the viscous force with respect to the surface tension force. As a result we take $Ca$ as the perturbation parameter in our asymptotic analysis. (vi) The radius of the droplet is assumed to be small enough as compared to the distance between the two plates of the micro channel. That is, the droplet dynamics remain unaffected due to the presence of any bounding walls and hence effectively it represents a droplet suspended in an unbounded medium. (vii) The solution for both flow field and temperature field are obtained under steady state assumption.

*3.2. Governing equations and boundary conditions*

The thermal problem, under the assumption of low thermal Péclet number, is governed by the Laplace equation for both the phases and can be expressed as

$$\left. \begin{array}{l} \bar{\nabla}^2 \bar{T}_i = 0, \\ \bar{\nabla}^2 \bar{T}_e = 0. \end{array} \right\} \quad (5)$$

The above set of governing equations are derived by neglecting the left hand side of the equation (3), thus ignoring any convective mode of heat transfer in the fluid.

We now look into the boundary conditions satisfied by the temperature field, $\bar{T}$. The far-field condition obeyed by the temperature outside the droplet is written below

$$\text{as } \bar{r} \to \infty, \quad \bar{T}_e = \bar{T}_\infty, \quad (6)$$

where $\bar{T}_\infty$ can be represented with respect to a cartesian coordinate system (origin attached to the bottom plate as shown in fig 1) as

$$\bar{T}_\infty = \bar{T}_L + \bar{G}\bar{y}, \quad (7)$$

where $\bar{T}_L$ is the reference temperature, which is the temperature of the lower plate and $\bar{G}$ is the constant temperature gradient acting along $\bar{y}$. For the present system we have considered the far field temperature to be linearly increasing in the cross-stream direction or the *y*-direction. Inside the droplet, the temperature $(\bar{T}_i)$ is bounded at the droplet center $(\bar{r}=0)$. At the droplet interface $(\bar{r}=\bar{r}_s=a\{1+g(\theta,\varphi)\})$, the temperature as well as the heat flux continuity is satisfied which can be written as

$$\begin{aligned}\text{at } \bar{r}=\bar{r}_s, \quad & \bar{T}_i=\bar{T}_e, \\ \text{at } \bar{r}=\bar{r}_s, \quad & k_i(\bar{\nabla}\bar{T}_i)\cdot\mathbf{n}=k_e(\bar{\nabla}\bar{T}_e)\cdot\mathbf{n}.\end{aligned} \quad (8)$$

Here $g(\theta,\varphi)$ is the shape correction due to deformation of the spherical droplet. $\mathbf{n}$ is the unit normal drawn on the droplet surface and is given by $\mathbf{n}=\bar{\nabla}\bar{F}/|\bar{\nabla}\bar{F}|$, where $\bar{F}=\bar{r}-\bar{r}_s$ is the equation of the surface of the droplet. The flow field, on the other hand, is governed by the Stokes equation as the Reynolds number is small and can be thus derived by neglecting the inertia terms on the left hand side of the generalized Navier-Stokes equation, that is equations (1) and (2). The continuity and the Stokes equation for either of the phases are given below

$$\begin{aligned}-\bar{\nabla}\bar{p}_i+\mu_i\bar{\nabla}^2\bar{\mathbf{u}}_i=\mathbf{0}, \quad & \bar{\nabla}\cdot\bar{\mathbf{u}}_i=0, \\ -\bar{\nabla}\bar{p}_e+\mu_e\bar{\nabla}^2\bar{\mathbf{u}}_e=\mathbf{0}, \quad & \bar{\nabla}\cdot\bar{\mathbf{u}}_e=0,\end{aligned} \quad (9)$$

The far field condition satisfied by both these fields are given below

$$\begin{aligned}\text{as } \bar{r}\to\infty, \quad & \bar{\mathbf{u}}_e=\bar{\mathbf{V}}_\infty-\bar{\mathbf{U}}, \\ \text{as } \bar{r}\to\infty, \quad & \bar{p}_e=\bar{p}_\infty,\end{aligned} \quad (10)$$

where $\bar{\mathbf{V}}_\infty$ is the imposed velocity field, $\bar{\mathbf{U}}$ is the droplet migration velocity with respect to the laboratory reference frame and $\bar{p}_\infty$ is the corresponding pressure field. The expression of the imposed velocity field $(\bar{\mathbf{V}}_\infty)$, with respect to a Cartesian coordinate system $(\bar{X},\bar{Y},\bar{Z})$ attached at the centroid of the undeformed droplet is given below

$$\begin{aligned}\bar{\mathbf{V}}_\infty &= \bar{V}_c\left\{k_0+k_1\bar{Y}+k_2\bar{Y}^2+k_3\bar{X}^2\right\}\hat{\mathbf{e}}_{\bar{X}}, \\ \text{where,} & \\ k_0 &= \frac{4\bar{y}_d}{\bar{H}}\left(1-\frac{\bar{y}_d}{\bar{H}}\right), \; k_1=\frac{4}{\bar{H}}\left(1-\frac{2\bar{y}_d}{\bar{H}}\right), \; k_2=-\frac{4}{\bar{H}^2}, \; k_3=0,\end{aligned} \quad (11)$$

where $\bar{H}$ is the channel height. The velocity and the pressure fields inside the droplet $(\bar{\mathbf{u}}_i, \bar{p}_i)$ are both bounded at the center of the droplet $(\bar{r} = 0)$. The boundary conditions for the flow field at the interface of the droplet $(\bar{r} = \bar{r}_s)$ are given by

$$\left.\begin{aligned} \text{at } \bar{r} = \bar{r}_s, \quad & \bar{\mathbf{u}}_i = \bar{\mathbf{u}}_e, \\ \text{at } \bar{r} = \bar{r}_s, \quad & \bar{\mathbf{u}}_i \cdot \mathbf{n} = \bar{\mathbf{u}}_e \cdot \mathbf{n} = 0, \\ \text{at } \bar{r} = \bar{r}_s, \quad & (\bar{\boldsymbol{\tau}}_e \cdot \mathbf{n} - \bar{\boldsymbol{\tau}}_i \cdot \mathbf{n}) = -\bar{\nabla}_s \bar{\sigma} + \bar{\sigma}(\bar{\nabla} \cdot \mathbf{n})\mathbf{n}, \end{aligned}\right\} \quad (12)$$

where $\bar{\nabla}_s = (\mathbf{I} - \mathbf{nn}) \cdot \bar{\nabla}$ is the surface gradient operator, and $\bar{\boldsymbol{\tau}}_i = -\bar{p}_i \mathbf{I} + \mu_i \left[\bar{\nabla}\bar{\mathbf{u}}_i + (\bar{\nabla}\bar{\mathbf{u}}_i)^T\right]$ and $\bar{\boldsymbol{\tau}}_e = -\bar{p}_e \mathbf{I} + \mu_e \left[\bar{\nabla}\bar{\mathbf{u}}_e + (\bar{\nabla}\bar{\mathbf{u}}_e)^T\right]$ represents the hydrodynamic stress comprising of hydrostatic and deviatoric stress components. The first boundary condition represents the continuity of the velocity field at the droplet interface. The second condition represents the no-penetration or the kinematic boundary condition. The third boundary condition indicates the stress balance at the droplet interface.

The equation of state indicating the linear dependence of surface tension on the interfacial temperature distribution is given by (Kim & Subramanian 1989; Das et al. 2017)

$$\bar{\sigma} = \bar{\sigma}_{eq} - \beta(\bar{T}_s - \bar{T}_L), \quad (13)$$

where $\bar{\sigma}_{eq}$ is the surface tension at the reference temperature $\bar{T}_L$, $\bar{T}_s = \bar{T}\big|_{\bar{r}=\bar{r}_s}$ is the interfacial temperature and $\beta = -d\bar{\sigma}/d\bar{T}_s$ is the temperature coefficient of surface tension.

Now we derive the non-dimensional set of governing equations and boundary conditions. We take the help of the following non-dimensional scheme: $r = \bar{r}/a$, $\mathbf{u} = \bar{\mathbf{u}}/V_c$, $T = (\bar{T} - \bar{T}_L)/|\bar{G}|a$, $\sigma = \bar{\sigma}/\bar{\sigma}_{eq}$, $p = \bar{p}/(\mu_e V_c/a)$, $\boldsymbol{\tau} = \bar{\boldsymbol{\tau}}/(\mu_e V_c/a)$. The non-dimensional numbers encountered are the thermal Marangoni number $Ma_T = \beta|\bar{G}|a/\mu_e V_c$ (signifying the relative strength of the thermally induced Marangoni stress with respect to the viscous stress) and the capillary number, $Ca = \mu_e V_c/\bar{\sigma}_{eq}$ (indicating the ratio of the viscous force to the surface tension force acting on the droplet). The different property ratios that appear in the analysis are the conductivity ratio, $\xi = k_i/k_e$, and the viscosity ratio, $\lambda = \mu_i/\mu_e$.

Using the above mentioned non-dimensional scales, we obtain the dimensionless governing equation for the temperature field as

$$\left.\begin{aligned}\nabla^2 T_i &= 0, \\ \nabla^2 T_e &= 0,\end{aligned}\right\} \tag{14}$$

which can be solved with the help of the following non-dimensional boundary conditions

$$\left.\begin{aligned}&\text{as } r \to \infty, T_e = \zeta r \sin\varphi P_{1,1}(\cos\theta), \\ &T_i \text{ is bounded at } r = 0, \\ &\text{at } r = r_s, \quad T_i = T_e, \\ &\text{at } r = r_s, \quad \xi(\nabla T_i)\cdot\mathbf{n} = (\nabla T_e)\cdot\mathbf{n}.\end{aligned}\right\} \tag{15}$$

The factor $\zeta = \bar{G}/|\bar{G}|$ indicates whether the temperature increases in the positive $y$-direction or in the negative $y$-direction, transverse to the imposed flow field. If $\zeta = 1$ the temperature increases linearly from the lower plate to the upper plate, whereas for $\zeta = -1$ there is a linear decrease in temperature in the positive $y$-direction. The non-dimensional governing equations for the flow field is given by

$$\left.\begin{aligned}-\nabla p_i + \lambda \nabla^2 \mathbf{u}_i &= \mathbf{0}, \quad \nabla\cdot\mathbf{u}_i = 0, \\ -\nabla p_e + \nabla^2 \mathbf{u}_e &= \mathbf{0}, \quad \nabla\cdot\mathbf{u}_e = 0.\end{aligned}\right\} \tag{16}$$

The corresponding boundary conditions for solving the above governing equations are as follows

$$\left.\begin{aligned}&\text{at } r \to \infty, \ (\mathbf{u}_e, p_e) = (\mathbf{V}_\infty - \mathbf{U}, p_\infty), \\ &\mathbf{u}_i \text{ is bounded at } r = 0, \\ &\text{at } r = r_s, \ \mathbf{u}_i\cdot\mathbf{n} = \mathbf{u}_e\cdot\mathbf{n} = 0, \\ &\text{at } r = r_s, \ \mathbf{u}_i = \mathbf{u}_e, \\ &\text{at } r = r_s, \ (\boldsymbol{\tau}_e\cdot\mathbf{n} - \boldsymbol{\tau}_i\cdot\mathbf{n}) = Ma_T \nabla_s T + \frac{\sigma}{Ca}(\nabla\cdot\mathbf{n})\mathbf{n}.\end{aligned}\right\} \tag{17}$$

The last boundary condition is the interfacial stress balance condition. Here we have made use of the non-dimensional form of the equation of state relating the surface tension $(\sigma)$ to the surface temperature $(T_s)$ in the following manner

$$\sigma = 1 - Ca(Ma_T T_s). \tag{18}$$

### 3.3. Spherical harmonic representation of field variables

As the governing equation for the temperature field is a Laplace equation (refer to equation (14)), the general solution for the temperature distribution for either of the phases can be written in the following form

$$\left.\begin{aligned}
T_i &= \sum_{n=0}^{\infty}\sum_{m=0}^{n}\left[a_{n,m}r^n\cos(m\varphi)+\hat{a}_{n,m}r^n\sin(m\varphi)\right]P_{n,m}(\cos\theta), \\
T_e &= \zeta r\sin\varphi P_{1,1}(\cos\theta)+\sum_{n=0}^{\infty}\sum_{m=0}^{n}\left[b_{-n-1,m}r^{-n-1}\cos(m\varphi)+\hat{b}_{-n-1,m}r^{-n-1}\sin(m\varphi)\right]P_{n,m}(\cos\theta),
\end{aligned}\right\} \quad (19)$$

where $P_{n,m}$ is an associate Legendre polynomial of order $m$ and degree $n$. The unknown constants present in the above expression $\left(a_{n,m},\hat{a}_{n,m},b_{-n-1,m},\text{and }\hat{b}_{-n-1,m}\right)$ can be determined by using the continuity of temperature and heat flux boundary conditions. The surface temperature can be expressed as

$$T_s = \sum_{n=0}^{\infty}\sum_{m=0}^{n}\left[T_{n,m}\cos(m\varphi)+\hat{T}_{n,m}\sin(m\varphi)\right]P_{n,m}(\cos\theta), \quad (20)$$

where the constant coefficients, $T_{n,m}$ and $\hat{T}_{n,m}$ represent surface spherical harmonics.

The general expression for the velocity and the pressure fields inside the droplet $(\mathbf{u}_i, p_i)$, which obey the Stokes equation as well as the continuity equation, can be written in terms of growing spherical harmonics by using the Lamb's general solution as (Hetsroni & Haber 1970)

$$\left.\begin{aligned}
\mathbf{u}_i &= \sum_{n=1}^{\infty}\left[\nabla\times(\mathbf{r}\chi_n)+\nabla\Phi_n+\frac{n+3}{2(n+1)(2n+3)\lambda}r^2\nabla p_n-\frac{n}{(n+1)(2n+3)\lambda}\mathbf{r}p_n\right], \\
p_i &= \sum_{n=0}^{\infty}p_n,
\end{aligned}\right\} \quad (21)$$

where $p_n$, $\Phi_n$ and $\chi_n$ are growing solid spherical harmonics of degree $n$, which can be expressed as

$$\left.\begin{aligned}
p_n &= \lambda r^n\sum_{m=0}^{n}\left[A_{n,m}\cos(m\varphi)+\hat{A}_{n,m}\sin(m\varphi)\right]P_{n,m}(\cos\theta), \\
\Phi_n &= r^n\sum_{m=0}^{n}\left[B_{n,m}\cos(m\varphi)+\hat{B}_{n,m}\sin(m\varphi)\right]P_{n,m}(\cos\theta), \\
\chi_n &= r^n\sum_{m=0}^{n}\left[C_{n,m}\cos(m\varphi)+\hat{C}_{n,m}\sin(m\varphi)\right]P_{n,m}(\cos\theta).
\end{aligned}\right\} \quad (22)$$

In a similar manner, using the Lamb's solution the velocity and pressure field outside the droplet can be expressed as (Hetsroni & Haber 1970)

$$\begin{aligned}\mathbf{u}_e &= (\mathbf{V}_\infty - \mathbf{U}) + \sum_{n=1}^{\infty}\left[\nabla\times(\mathbf{r}\chi_{-n-1}) + \nabla\Phi_{-n-1} - \frac{n-2}{2n(2n-1)}r^2\nabla p_{-n-1} + \frac{n+1}{n(2n-1)}\mathbf{r}p_{-n-1}\right], \\ p_e &= p_\infty + \sum_{n=0}^{\infty} p_{-n-1},\end{aligned} \quad (23)$$

where $(\mathbf{V}_\infty, p_\infty)$ are the velocity and pressure fields as $r \to \infty$ and $p_{-n-1}, \Phi_{-n-1}, \chi_{-n-1}$ represent the decaying solid spherical harmonics of the following form

$$\begin{aligned}p_{-n-1} &= r^{-n-1}\sum_{m=0}^{n}\left[A_{-n-1,m}\cos(m\varphi) + \hat{A}_{-n-1,m}\sin(m\varphi)\right]P_{n,m}(\cos\theta), \\ \Phi_{-n-1} &= r^{-n-1}\sum_{m=0}^{n}\left[B_{-n-1,m}\cos(m\varphi) + \hat{B}_{-n-1,m}\sin(m\varphi)\right]P_{n,m}(\cos\theta), \\ \chi_{-n-1} &= r^{-n-1}\sum_{m=0}^{n}\left[C_{-n-1,m}\cos(m\varphi) + \hat{C}_{-n-1,m}\sin(m\varphi)\right]P_{n,m}(\cos\theta).\end{aligned} \quad (24)$$

The constant coefficients $A_{n,m}$, $B_{n,m}$, $C_{n,m}$, $A_{-n-1,m}$, $B_{-n-1,m}$, $C_{-n-1,m}$, $\hat{A}_{n,m}$, $\hat{B}_{n,m}$, $\hat{C}_{n,m}$, $\hat{A}_{-n-1,m}$, $\hat{B}_{-n-1,m}$ and $\hat{C}_{-n-1,m}$ in the equations (22) and (24) can be obtained by substituting equations (21) and (23) in the boundary conditions at the interface of the droplet (the kinematic boundary conditions, the tangential velocity continuity and the tangential stress balance conditions) and then solving them.

*3.4. Asymptotic solution*

We now proceed towards solving the governing equations for flow field as well as the temperature field with the help of the boundary conditions stated above. Even with all the assumptions made, the simplified governing equations along with the boundary conditions cannot be solved analytically as the flow field is still coupled with the temperature field through the equation of state (18) and the shear stress balance. This makes the system of differential equations non-linear. Towards tackling this non-linearity, we use the regular perturbation method taking the capillary number as the perturbation parameter. This is possible only for small deformation of the droplet. So any generic variable, $\psi$, can be expanded in a power series as follows

$$\psi = \psi^{(0)} + Ca\,\psi^{(Ca)} + O(Ca^2), \quad (25)$$

where $\psi^{(0)}$ represents the leading order term corresponding to $Ca = 0$ and $\psi^{(Ca)}$ is $O(Ca)$ shape correction to the solution. Considering $g^{(Ca)}$ and $g^{(Ca^2)}$ as the $O(Ca)$ and $O(Ca^2)$ correction to the spherical shape of the droplet, the deformed droplet can be represented as follows

$$r_s = 1 + Ca\, g^{(Ca)}(\theta,\varphi) + Ca^2 g^{(Ca^2)}(\theta,\varphi) + O(Ca^3). \tag{26}$$

### 3.4.1. Leading Order solution

The leading order solution is obtained for an undeformed droplet that is for $Ca = 0$. We first solve the temperature field inside as well as outside the droplet from the leading order boundary conditions obtained from equation (15). The expression for the leading order temperature field for both the phases is given below

$$\left. \begin{array}{l} T_e^{(0)} = r\zeta \left\{ 1 + \dfrac{1}{r^3}\left(\dfrac{1-\xi}{2+\xi}\right) \right\} \cos\varphi P_{1,1}(\cos\theta), \\[6pt] T_i^{(0)} = r\zeta\left(\dfrac{3}{\xi+2}\right)\cos\varphi P_{1,1}(\cos\theta). \end{array} \right\} \tag{27}$$

The surface temperature thus obtained from above is given by $T_s^{(0)} = 3\zeta/(\xi+2)\cos\varphi P_{1,1}(\cos\theta)$ where the only non zero coefficient of the surface harmonic $\cos\varphi P_{1,1}(\cos\theta)$ is given by $T_{1,1} = 3\zeta/(\xi+2)$.

The leading order solution for the velocity and pressure fields are obtained by substituting equations (21)-(24) in the boundary conditions for leading order. We then use the force-free condition to calculate the droplet migration velocity. This force-free condition is obtained by equating the net drag force on the droplet to zero which is given by

$$\mathbf{F}_D^{(0)} = 4\pi\nabla\left(r^3 p_{-2}^{(0)}\right), \tag{28}$$

where the decaying solid harmonic $p_{-2}$ is given by

$$p_{-2} = r^{-3}\left[ A_{-2,0}^{(0)} P_{2,0}(\cos\theta) + A_{-2,1}^{(0)} \cos\phi P_{2,1}(\cos\theta) + \hat{A}_{-2,1}^{(0)} \sin\phi P_{2,1}(\cos\theta) \right]. \tag{29}$$

The constant coefficients $A_{-2,0}^{(0)}$, $A_{-2,1}^{(0)}$ and $\hat{A}_{-2,1}^{(0)}$ which has been found previously from the boundary conditions are substituted in equation (29) to calculate $p_{-2}$ and hence the migration velocity is calculated from the condition $\mathbf{F}_D^{(0)} = 0$. Thus the expressions of the leading order axial velocity (along *x*-direction), $U_x^{(0)}$ and the cross-stream velocity (along *y*-direction) $U_y^{(0)}$, thus obtained, are provided below

$$U_x^{(0)} = -\frac{4\lambda}{H^2(3\lambda+2)} + 4\frac{y_d}{H}\left(1 - \frac{y_d}{H}\right),$$

$$U_y^{(0)} = \frac{2\zeta Ma_T}{(\xi+2)(3\lambda+2)}, \qquad (30)$$

$$U_z^{(0)} = 0.$$

It can be seen from the above expressions that for the leading order, the interfacial temperature distribution has no effect on the axial migration velocity, whereas the cross-stream migration velocity in the *y*-direction originates solely due to the transversely applied constant temperature gradient. We next calculate the $O(Ca)$ deformation with the help of the normal stress balance. The $O(Ca)$ correction to the shape of the droplet is found out by using orthogonality condition for associate Legendre polynomials in the normal stress balance and is given by

$$g^{(Ca)} = \sum_{n=2}^{\infty}\sum_{m=0}^{n}\left[L_{n,m}^{(Ca)}\cos(m\varphi) + \hat{L}_{n,m}^{(Ca)}\sin(m\varphi)\right]P_{n,m}(\cos\theta), \qquad (31)$$

where $L_{n,m}^{(Ca)}$ and $\hat{L}_{n,m}^{(Ca)}$ are constant coefficients given by

$$\left.\begin{aligned}L_{3,0}^{(Ca)} &= \frac{1}{10}\frac{11\lambda+10}{(1+\lambda)H^2},\\ L_{2,1}^{(Ca)} &= \frac{1}{6}\frac{(H-2y_d)(19\lambda+16)}{(1+\lambda)H^2},\\ L_{3,2}^{(Ca)} &= -\frac{1}{60}\frac{11\lambda+10}{(1+\lambda)H^2}.\end{aligned}\right\} \qquad (32)$$

### 3.4.2. $O(Ca)$ solution

Using the above results we move forward towards calculating the $O(Ca)$ temperature, velocity and pressure fields. We first calculate the $O(Ca)$ temperature field from the boundary conditions based on the deformed surface of the droplet and derived from equation (8). The expression of the temperature field thus obtained for this order is given by

$$\left.\begin{aligned}T_i^{(Ca)} &= \zeta\left[ra_{1,0}^{(Ca)}P_{1,0} + r^2 a_{2,1}^{(Ca)}\cos\varphi P_{2,1} + r^3 a_{3,0}^{(Ca)}P_{3,0} + r^3 a_{3,2}^{(Ca)}\cos(2\varphi)P_{3,2}\right],\\ T_e^{(Ca)} &= \zeta\left[r^{-2}b_{-2,0}^{(Ca)}P_{1,0} + r^{-3}b_{-3,1}^{(Ca)}\cos\varphi P_{2,1} + r^{-4}b_{-4,0}^{(Ca)}P_{3,0} + r^{-4}b_{-4,2}^{(Ca)}\cos(2\varphi)P_{3,2}\right],\end{aligned}\right\} \quad (33)$$

where the constant coefficients present in the above equation $a_{1,0}^{(Ca)}$, $a_{2,1}^{(Ca)}$, $a_{3,0}^{(Ca)}$, $a_{3,2}^{(Ca)}$, $b_{-2,0}^{(Ca)}$, $b_{-3,1}^{(Ca)}$, $b_{-4,0}^{(Ca)}$ and $b_{-4,2}^{(Ca)}$ are given in Appendix A.

Next the $O(Ca)$ solution for the flow field is found out by deriving the boundary conditions on the deformed surface of the droplet from equation (17) and then solving them (except the normal stress boundary condition) simultaneously. The droplet migration velocity for this order comes out to be

$$\left.\begin{aligned}U_x^{(Ca)} &= \zeta\left(\frac{H-2y_d}{5H^2}\right)\left\{\frac{(19\lambda+16)(3\lambda\xi-21\lambda+17\xi+16)}{(\xi+2)^2(2+3\lambda)^2(\lambda+1)}\right\}Ma_T,\ U_z^{(Ca)} = 0\\ U_y^{(Ca)} &= \frac{8}{105}\left(\frac{H-2y_d}{H^4}\right)\left[\frac{198\lambda^5 - 1242\lambda^4 - 7327\lambda^3 - 6292\lambda^2 + 1843\lambda + 2320}{(\lambda+1)^2(2+3\lambda)^2(\lambda+4)}\right].\end{aligned}\right\} \quad (34)$$

It can be seen from the above expressions that there is no contribution of the imposed temperature field on the cross-stream migration velocity for this order. However this is not the case for the axial velocity $\left(U_x^{(Ca)}\right)$, where there is seen to be a direct dependence on the imposed temperature field.

## 4. Numerical approach

### 4.1. Numerical setup

We first show a schematic of the numerical setup in figure 2, where we have a Newtonian droplet suspended in another Newtonian fluid. The droplet is initially located at a distance $\bar{y}_d$ from the lower plate. The quantity $\Phi$ is called the level set variable. It is constant at the interface and varies between 0.5 and 1 inside the droplet whereas between 0.5 and 1 in the suspending phase. The upper plate is maintained at a higher temperature that the lower plate with the distance between the two plates being $\bar{H}$. The temperature is assumed to vary linearly in the transverse direction. The upper and lower boundaries (walls) of the computational domain are denoted by $S_T$ and $S_B$ respectively. A pressure driven flow is imposed in the micro channel at $S_L$.

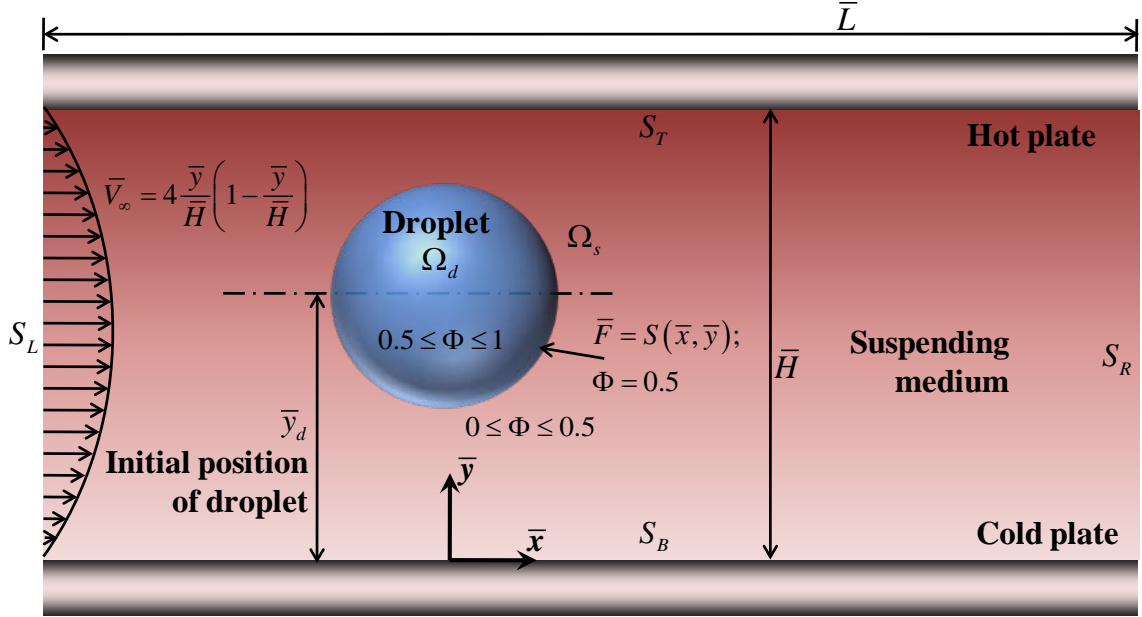

FIGURE 2. Schematic of the numerical setup consisting of a viscous Newtonian droplet immersed in a Newtonian carrier fluid within a parallel plate channel. The 2D computational domain considered has a length and width of $\bar{L}$ and $\bar{H}$ respectively. The droplet domain is denoted by $\Omega_d$ and the carrier phase domain by $\Omega_s$. The left and right boundaries $(S_L, S_R)$ of the computational domain are periodically connected. The droplet has a radius of $a$ and is placed at a position $\bar{y}_d$ from the bottom plate.

## 4.2. Governing equations and boundary conditions

We use the conservative level set (CLS) approach to solve the two phase flow. Details of this method are been stated in various literatures (Sethian & Smereka 2003; Nguyen & J.-C. Chen 2010; Balcázar et al. 2016; Olsson & Kreiss 2005). Here we provide a brief overview of the methodology and the relevant governing equations to be solved. The entire domain consists of a droplet and a carrier phase sub-domain, $(\Omega_d, \Omega_s)$ as shown in figure 2. As per Olsson and Kreiss, (2005) the level set function, $\Phi$, varies smoothly from 0 to 1 across the interface. The droplet interface is located at $\Phi = 0.5$. The level set function, $\Phi$, is assumed to vary in either of the phases in the following manner: $\Omega_d$: $0.5 \leq \Phi \leq 1$ and $\Omega_s$: $0 \leq \Phi \leq 0.5$. All the governing equations as well as the corresponding boundary conditions are written in their dimensionless form using the dimensionless scheme mentioned earlier in section 3.2.

Here we show the governing equations and boundary conditions that are used for the purpose of a full scale numerical simulation. Thus we don't consider any of the assumptions that were made in the previous section. The governing equation for $\Phi$, which signifies the reinitialized convection of the interface is given by (Balcázar et al. 2016; Wang et al. 2008; Olsson & Kreiss 2005; Nguyen & Chen 2010)

$$\frac{\partial \Phi}{\partial t} + \mathbf{u} \cdot \nabla \Phi = \gamma \nabla \cdot \left[ \varepsilon \nabla \Phi - \Phi(1-\Phi) \frac{\nabla \Phi}{|\nabla \Phi|} \right], \qquad (35)$$

where $\gamma \left(= \bar{\gamma}/\bar{V_c}\right)$ is the re-initialization parameter and $\varepsilon \left(= \bar{\varepsilon}/a\right)$ is the parameter governing interface thickness. The different fluid properties are expressed in terms of the level set function as follows

$$\left.\begin{array}{l} k = \Phi k_i + (1-\Phi) k_e, \\ \mu = \Phi \mu_i + (1-\Phi) \mu_e. \end{array}\right\} \qquad (36)$$

In the numerical simulations, as we have considered a neutrally buoyant droplet, we can write $\rho = \rho_i = \rho_e$. The non-dimensional continuity and the Navier-Stokes equation is obtained by using the same non-dimensional scheme (as was used in the previous section) in equations (1) and (2). Further substituting (36) into the dimensionless governing equations, we finally get

$$\nabla \cdot \mathbf{u} = 0, \qquad (37)$$

$$Re\{\omega\Phi + (1-\Phi)\}\left\{\frac{\partial \mathbf{u}}{\partial t} + \nabla \cdot (\mathbf{uu})\right\} = -\nabla p + \nabla \cdot \left[\{\lambda\Phi + (1-\Phi)\}\{\nabla \mathbf{u} + (\nabla \mathbf{u})^T\}\right] + \mathbf{F}_{st} + \mathbf{F}_v, \quad (38)$$

where $\text{Re} = \rho_e \bar{V_c} H / \mu_e$ is the Reynolds number signifying the ratio the inertia and viscous forces, $\omega = \rho_i / \rho_e$ is the ratio of the density of the inner fluid to that of the outer fluid ($\omega = 1$ for a neutrally buoyant droplet). We have considered the body forces in the governing equations to take into account the effect of the viscous as well as the surface tension forces on the droplet dynamics. $\mathbf{F}_v = \left(8a^2/H^2\right)\hat{\mathbf{x}}$ is the dimensionless volumetric force acting on the fluid due to the imposed pressure gradient, and $\mathbf{F}_{st} = \left(\frac{\sigma}{Ca}\right)\kappa \delta H^2 \mathbf{n}$ is the non-dimensional surface tension force. Here $\mathbf{n} = \left.\frac{\nabla \Phi}{|\nabla \Phi|}\right|_{\Phi=0.5}$ is a unit vector representing the normal drawn on the droplet surface and $\kappa$ is the local interfacial curvature given by

$$\kappa = \nabla \cdot \mathbf{n}, \qquad (39)$$

$\delta = \delta(\Phi)$ is the Dirac delta function that ensures that $\mathbf{F}_{st}$ acts on the surface of the droplet, $F$.

The non-dimensional energy equation, for solving the temperature field, obtained from equation (3) and equation (36) is written as

$$Pe_T \{\omega\Phi + (1-\Phi)\} \left(\frac{\partial T}{\partial t} + \mathbf{u}\cdot\nabla T\right) = \nabla\cdot\left[\{\xi\Phi + (1-\Phi)\}\nabla T\right], \tag{40}$$

where $Pe_T = \bar{V}_c a/\alpha_e$ is the thermal Péclet number and $\alpha_e = k_e/\rho_e c_p$ is the thermal diffusivity of the carrier phase. The thermal conductivity is taken to be a constant in space and time. Also any viscous dissipation or radiation from either of the phases is neglected.

Now we state the dimensionless boundary conditions at the upper and the lower plates, $S_T$ and $S_B$ (see figure 2) required to solve above set of equations. We denote the unit vector normal to a wall (or plate) as $\mathbf{n}_s$.

$$\begin{aligned}
&\text{(i)} \quad \text{No slip}: \mathbf{u} - (\mathbf{u}\cdot\mathbf{n}_s)\mathbf{n}_s = \mathbf{0}, \\
&\text{(ii)} \quad \text{No penetration}: \mathbf{u}\cdot\mathbf{n}_s = 0, \\
&\text{(iii)} \quad \text{No flux}: \mathbf{n}_s \cdot \nabla\Phi = 0,
\end{aligned} \tag{41}$$

and finally the condition required for the periodicity of the velocity and pressure fields as well as for the level set function at $S_L$ and $S_R$ is given by

$$\begin{aligned}
&\text{(i)} \quad \mathbf{u}(\mathbf{x}) = \mathbf{u}(\mathbf{x}+L), \\
&\text{(ii)} \quad p(\mathbf{x}) = p(\mathbf{x}+L), \\
&\text{(iii)} \quad \Phi(\mathbf{x}) = \Phi(\mathbf{x}+L).
\end{aligned} \tag{42}$$

It can be seen that equations (35), (37), (38) and (40) are non-linear coupled partial differential equations, which cannot be solved analytically. So we use the conservative level set approach to solve the flow field and the temperature field numerically.

In accordance to the figure 2, we choose a 2D rectangular computational domain of aspect ratio $3\times 1$, tessellated with uniform grids. The equations (35), (37), (38) and (40) are solved along with the relevant boundary conditions, equations (41) and (42), using the finite element based conservative level set method with second order Lagrangian triangular elements. We take the help of commercially available software, COMSOL MULTIPHYSICS, to solve these equations. The initial value of the level set function, $\Phi$, is taken as

$$\begin{aligned}
&\text{carrier phase}: \Phi(t=0) = 0, \\
&\text{droplet phase}: \Phi(t=0) = 1.
\end{aligned} \tag{43}$$

The initial velocity profile of the flow field is that of a plane Poiseuille flow and is given by

$$\mathbf{u}(t=0) = \mathbf{V}_\infty = 4\left(\frac{y}{H}\right)\left(1-\frac{y}{H}\right)\mathbf{e}_x, \quad (44)$$

where $\mathbf{e}_x$ is the unit vector in the *x*-direction. The initial temperature of the upper and the lower plates are held constant and are taken to be $T_L = y_{min}$ and $T_H = y_{min} + H$ respectively, where $y_{min}$ is the position of the lower plate in the computational domain.

## 5. Results and discussions

We first validate our numerical results with two of previously published works. In the first place, we match the temporal variation of the lateral position of the droplet with that as obtained by Mortazavi and Tryggvason (2000) in their work. Secondly we match the axial migration velocity of a droplet due to an imposed temperature gradient with that obtained by Nas and Tryggvason (2003). The details of these validations are provided in Appendix B. We also carry out a grid-independency check to ensure that our numerical results are not affected much by a change in the element size. The details of this check along with a plot showing the time variation of the lateral position of the droplet for different values of the maximum element size chosen are provided in Appendix C. We next compare our numerical as well as the analytical results obtained in the present analysis.

### *5.1. Comparison between numerical and analytical results*

In this section we provide a qualitative matching of the trend in the variation of the transverse position of the droplet with time. In figure 3(a) we have shown the temporal variation of the lateral position of the droplet for different initial positions $\left(y_d(t=0) = 2.35, 2.55\right)$, obtained numerically. The confinement ratio of the system, which is the ratio of the droplet radius $(a)$ to the channel width $(\bar{H})$, has been taken as 0.25, in order to obtain a good match with the theoretical results. The other parameters used in this simulation are $Re = 0.01$, $Ca = 0.2$, $Ma_T = 0.1$, $\lambda = 1$ and $\xi = 1$. The corresponding theoretical plot has been shown in figure 3(b). We obtain this temporal variation of lateral position of the droplet by substituting $U_y = dy_d/dt$ in the expression for the cross-stream migration velocity of the droplet, which is given by

$$U_y = dy_d/dt = U_y^{(0)} + Ca\, U_y^{(Ca)} + O(Ca^2), \quad (45)$$

where the expression for $U_y^{(0)}$ and $U_y^{(Ca)}$ are given in equations (30) and (34) respectively. Thus solving for $y_d(t)$ from the above equation we get

$$y_d(t) = \frac{\varpi}{\upsilon} + \left\{ y_d(t=0) - \frac{\varpi}{\upsilon} \right\} \exp\left(-\frac{t}{\tau_c}\right), \tag{46}$$

where expressions of the constants $\varpi$ and $\upsilon$ are given in Appendix D. $\tau_c (=1/\upsilon)$ in the above expression represents the characteristic time constant. In the figure 3 we have taken $\lambda = 0.1$ and $Ma_T = 0.015$. As can be seen there is a good qualitative match between the two plots.

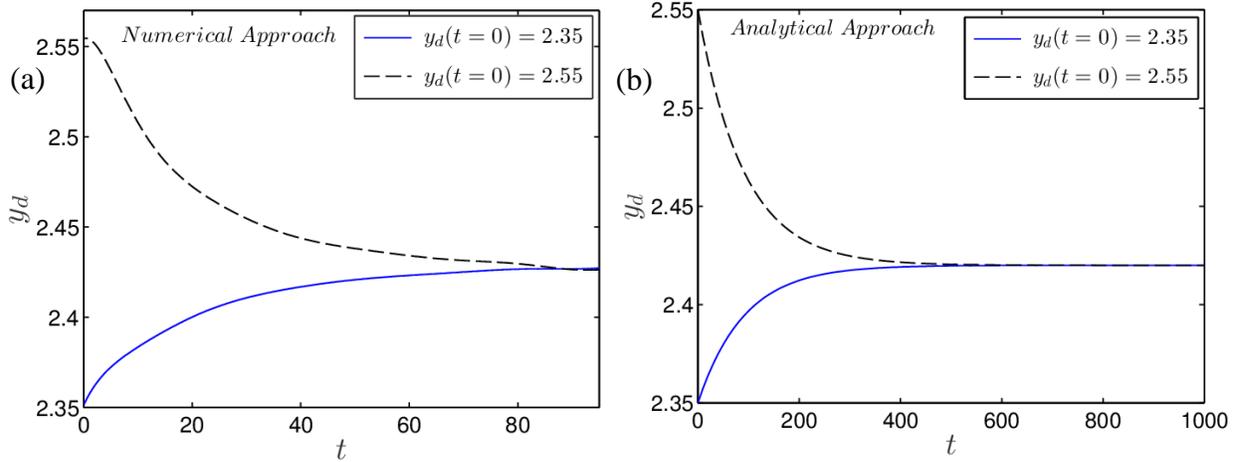

FIGURE 3. Variation of the lateral position of the droplet with time for different initial positions, $y_d(t=0) = 2.35, 2.55$. (a) Numerically obtained result using conservative level set method. The channel dimensions are $4 \times 12$. The other non-dimensional parameters are $Ma_T = 0.1$, $\lambda = 1$, $Pe_T = 0.1$, and $Re = 0.01$. (b) Analytically obtained result. The different constant parameter values are: $Ca = 0.2$, $\omega = \xi = 1$, $\lambda = 0.1$, $Ma_T = 0.015$ and $H = 4$.

Next we have also shown a qualitative match of the variation of the transverse position of the droplet with time, for different values of $Ma_T (0.001, 0.01, 0.1, 1)$, in figure 4. In figure 4(a) we have shown the numerical results, with the other parameters being the same as in figure 3(a). A theoretical plot obtained theoretically for the same values of $Ma_T$ is shown in figure 4(b). Although we don't have an exact match, but the trend of cross-stream migration of the droplet remains the same. For instance, in both of the cases the droplet migrates towards the channel centerline and reaches a steady state subsequently, provided $Ma_T < 0.1$. However, for $Ma_T = 0.1$ or $1$, the droplet is always found to migrate away from the channel centerline.

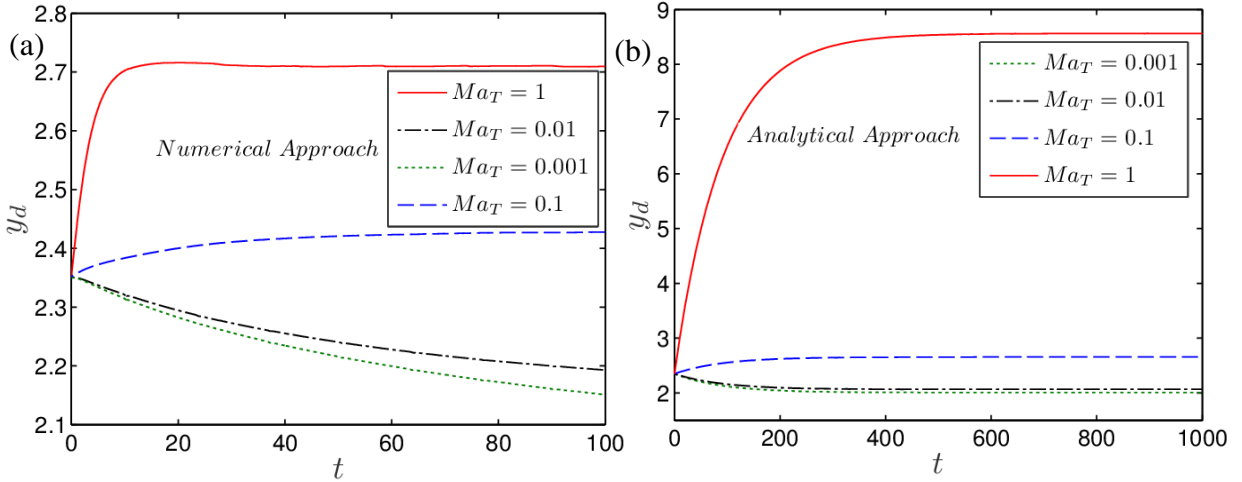

FIGURE 4. Temporal variation of the lateral position of the droplet for different values of Marangoni number, $Ma_T = 0.001, 0.01, 0.1, 1$. (a) Numerically obtained result using conservative level set method. The channel dimensions are $4 \times 12$. The other parameters are $\lambda = 1$, $Pe_T = 0.1$, and $Re = 0.01$. (b) Analytically obtained result. The different parameters involved are: $Ca = 0.2$, $\omega = \xi = 1$, $\lambda = 0.1$, and $H = 4$.

In the 2D numerical simulation carried out the effect of bounding wall, thermal convection as well as finite deformation of the droplet have been considered. As $H = 4$ in the above numerical simulations, wall effects and associated hydrodynamic lift on the droplet is not negligible. On the contrary the theoretical analysis is done in 3D for a droplet suspended in an unbounded domain with negligible effect of thermal convection. A full scale 3D simulation is not possible as the computational cost is too high to afford. Hence a simplification is done in our analysis by using a 2D computational domain to simulate a 3D problem. This simplification has been previously justified by Mortazavi and Tryggvason (2000) and Stan et al. (2011) where they have compared the numerical results obtained from 2D and 3D simulations and found similar trend for either of the cases. A proper one-to-one matching of our theoretical and numerical results is thus not possible, however, a qualitative match in the trend of temporal variation of transverse position of the droplet is found.

### 5.2. Effect of $Ca$ and $Ma_T$ on cross-stream migration

We now explore the variation of the transverse position of the droplet with time. We run full scale simulations for different values of $Ca$ and $Ma_T$. In figure 5 we show the effect of $Ma_T$ on the cross-stream migration of a droplet for a constant value of $Ca$. The other parameters used in these simulations are $Re = 4$, $Pe_T = 20$, $H = 3$ and $L = 9$. It can be observed from figure 5 that increase in $Ma_T$ and hence increase in the thermal Marangoni stress shifts the steady state position of the droplet further away from the channel centerline. As the droplet is placed at an off center position with respect to the centerline, surface velocity is different at different positions on its surface: higher at the bottom hemisphere and lower at the upper hemisphere, when the droplet is placed above the centerline. As a result, in the absence of any temperature gradient in the far field, a hydrodynamic force acts on the droplet which forces it to always migrate towards the centerline of the channel.

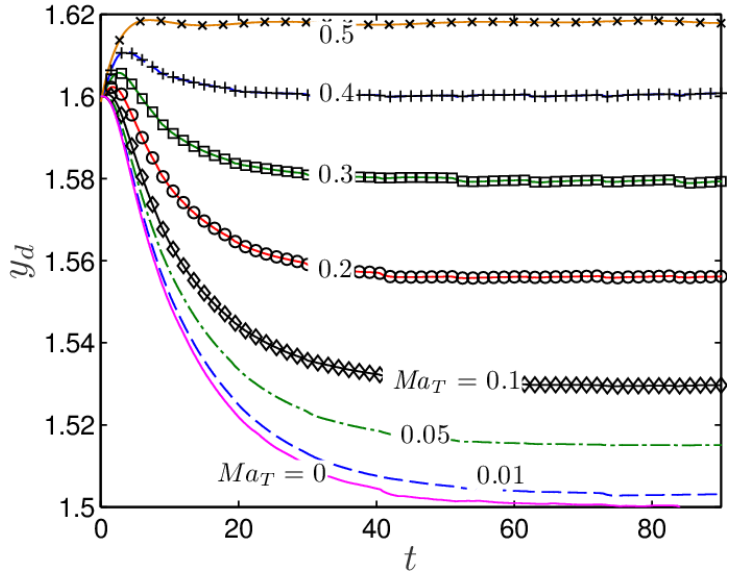

FIGURE 5. Temporal evolution of the transverse position of the droplet for different values of $Ma_T$. The other parameter values which are maintained constant throughout these simulations are: $Ca = 0.3$, $Re = 4$, $\omega = \xi = \lambda = 1$, $Pe_T = 20$, $H = 3$, $L = 9$.

However, in the presence of a transverse temperature gradient, as shown in figure 2, the droplet may migrate towards or away from the centerline depending on the strength of the temperature field. The strength of this temperature field is manifested by the thermal Marangoni stress or the tangential stress jump at the droplet interface. This Marangoni stress acting on the droplet opposes the hydrodynamic force due to the imposed flow. When the thermal Marangoni stress overcomes this hydrodynamic force, the droplet migrates away from the centerline of the channel.

In figure 6, we show the effect of $Ca$ on the lateral migration of the droplet, for a constant value of $Ma_T$. It is seen that higher the value of $Ca$, higher is the effect of the hydrodynamic force due to imposed flow. A higher value of $Ca$ signifies that the surface tension force driving the flow is lower as compared to the viscous force due to the imposed flow which indicates lower thermal Marangoni stress. Thus for higher values of $Ca$, the thermocapillary migration of the droplet is dominated by the imposed Poiseuille flow, and as a result the droplet moves towards the channel centerline although it might not reach the channel axis (see figure 6) at steady state. It can be seen from figure 6 that the droplet migrates towards the upper plate of the channel for $Ma_T = 0.4$ and $Ca = 0.1$. However, for $Ca = 0.5$, the droplet migrates towards the channel centerline for the same value of $Ma_T$.

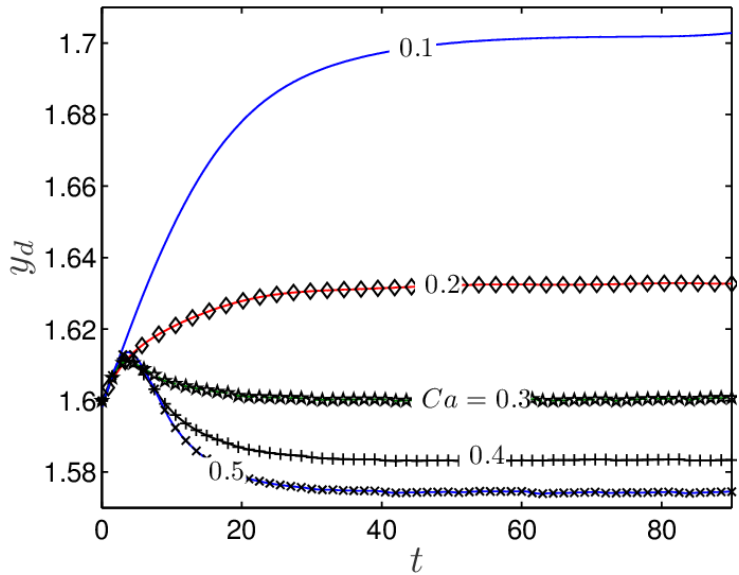

FIGURE 6. Temporal variation of the lateral position of the droplet for different values of $Ca$. The other parameter values used are: $Ma_T = 0.4$, $Re = 4$, $\omega = \xi = \lambda = 1$, $Pe_T = 20$, $H = 3$ and $L = 9$. The droplet is initially positioned above the channel centerline with $y_d(t=0) = 1.6$.

Next we construct a regime diagram to show the droplet trajectory for different values of $Ca$ and $Ma_T$. figure 7 shows two distinguishable regimes: one of them indicates that the droplet migrates towards the centerline of the channel and the other signifies that the droplet moves towards the channel wall. The 'red square' data points indicate the values of $(Ca, Ma_T)$ for the simulations which show that the droplet migrate towards the channel centerline while the 'blue circle' data points represent the data points where the droplet migrate towards the wall (upper plate for the present case). The 'green' region signifies the regime where the droplet migrates towards the wall, and the 'pink' region indicates the regime where the droplet migrates towards the channel centerline. The borderline separating the two regimes represents the magnitude of the critical Marangoni number, which is defined as the Marangoni number corresponding to which there is zero cross-stream migration velocity of the droplet. The magnitude of the critical thermal Marangoni stress (corresponding to the critical Marangoni number) is such that it exactly neutralizes the effect of hydrodynamic force due to the imposed flow and there is no migration of the droplet.

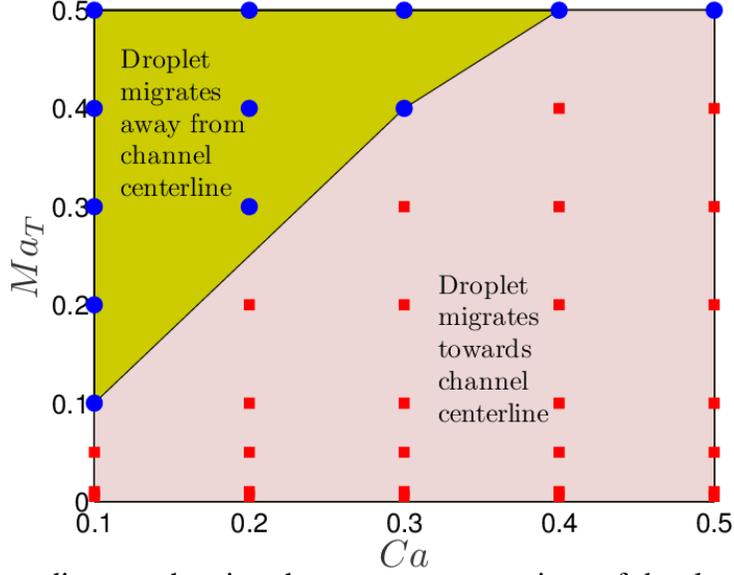

FIGURE 7. Regime diagram showing the two separate regions of droplet migration. The other parameters are $Re = 4$, $Pe_T = 20$, $H = 3$, and $L = 9$.

*5.3. Effect of initial droplet position, $y_d(t=0)$*

Now we explore the possible effects of change in initial position of the droplet on its lateral migration. We first show in figure 8(a), the temporal variation of the lateral position of the droplet for different initial positions of the droplet: above, below and on the centerline $(y_d(t=0)=1.4, 1.5, 1.6)$, when suspended in an isothermal Poiseuille flow field $(Ma_T = 0)$. We see that when the droplet is initially placed on the channel axis $(y_d = H/2)$, it migrates solely in the axial direction. However when the droplet is eccentrically placed $(y_d(t=0)=1.4$ or $1.6)$, it always migrates towards the centerline of the channel, and the steady state transverse position of the droplet becomes the channel centerline $(y_d(t=100) = H/2)$ itself.

In figure 8(b), we consider a constant temperature gradient in the transverse direction. Under the effect of a linearly varying temperature field, the steady state transverse position of the droplet no longer remains the channel centerline. Since the upper plate is the hotter one, the thermocapillary effect or the temperature induced Marangoni stress causes the droplet to migrate upwards. However, due to the presence of the imposed flow, the droplet doesn't follow a vertical path. Depending on the initial position of the droplet as well as the dominating factor out of these two: the hydrodynamic force acting on the droplet due to the imposed Poiseuille flow (governed by $Ca$) and the thermal Marangoni stress generated due to variation of the interfacial temperature (governed by $Ma_T$), the droplet may migrate towards or away from the channel centerline. Whatever may be the case, the droplet always migrates to the same steady state position irrespective of the initial position of the droplet, for any constant value of $Ca$ and $Ma_T$. This steady state position of the droplet may $(y_d = H/2)$ or may not $(y_d \neq H/2)$ be on the channel centerline depending on whether $Ma_T = 0$ or $Ma_T \neq 0$, respectively. For the plot shown in figure 8(b), we have $Ma_T = 0.1$ and $Ca = 0.3$, where the droplet reaches a steady state position, $y_d = 1.528$ which is above the channel centerline.

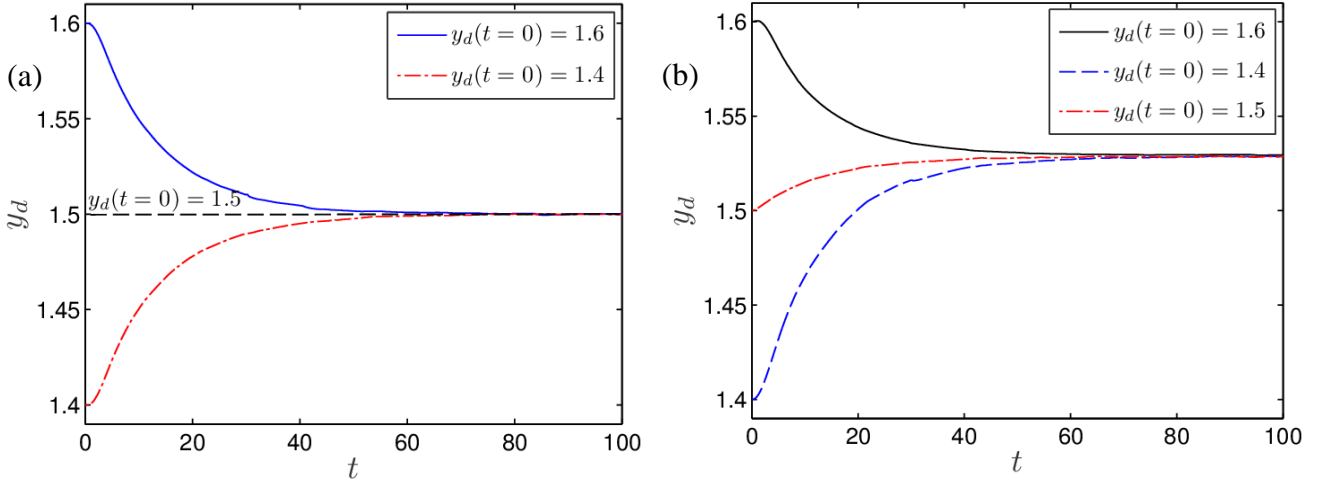

FIGURE 8. Temporal variation of the lateral position of the droplet for different initial positions, $y_d(t=0) = 1.4, 1.5, 1.6$ in (a) an isothermal Poiseuille flow field $(Ma_T = 0)$, (b) an non-isothermal Poiseuille flow field $(Ma_T = 0.1)$. The other parameters which are kept constant in these simulations are: $\omega = \xi = \lambda = 1$, $Re = 4$, $Pe_T = 20$, $a/\bar{H} = 0.33$, $Ca = 0.3$.

## 5.4. Effect of confinement ratio, $a/\bar{H}$

Now, we look into the effect of the confinement ratio on the droplet cross-stream migration. As seen from figure 9, the droplet migrates towards the channel centerline and reaches its steady state position faster for a system with higher confinement ratio. As the confinement ratio decreases, the time taken by the droplet to reach the steady state position increases. Also, the steady state position of the droplet is shifted further away from the channel centerline for a lower confinement ratio. On further reduction of the confinement ratio $(a/\bar{H} = 0.2857)$, the droplet migrates away from the channel centerline. A proper reasoning for this can be given as follows: when the confinement ratio is large or the droplet is larger as compared to the channel width, the difference in the surface velocity on the droplet interface between the upper and the lower hemisphere, due to the effect of imposed Poiseuille flow, is larger as compared to a system with low confinement ratio. As a result, a larger hydrodynamic force acts on the droplet for a system with higher value of $a/\bar{H}$. Thus the confinement ratio can be increased such that hydrodynamic force, acting on the droplet, dominates the effect of the thermal Marangoni stress and as a result the droplet migrates towards the center. This also explains the fact that a higher value of $a/H$ causes the droplet to reach its steady state position quickly and also more closer to the channel centerline.

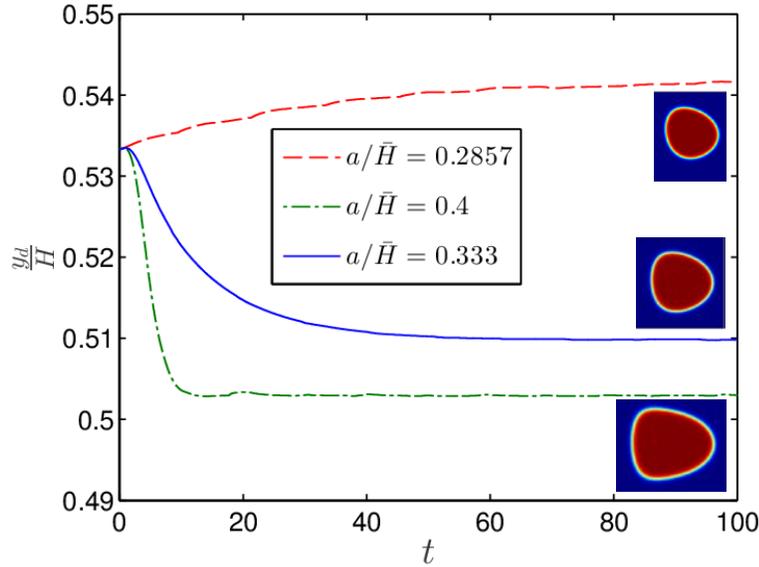

FIGURE 9. Variation of lateral position of droplet (normalized with channel height) with time for different values of confinement ratio $a/\bar{H} = 0.2857, 0.333, 0.4$. The droplet shapes for each confinement ratios at $t = 100$ are shown as insets. The other parameters are $\omega = \xi = \lambda = 1$, $Re = 4$, $Pe_T = 20$, $Ca = 0.3$, $Ma_T = 0.1$. The initial position of the droplet for all the three cases are taken as $y_d(t=0) = 1.6$.

When $a/\bar{H}$ is sufficiently low, the difference in surface velocity between the upper and the lower hemisphere of the droplet reduces. Hence the thermal Marangoni stress overcomes the effect of the hydrodynamic force due to the imposed flow, forcing the droplet to migrate towards the wall. We also show the steady state droplet shapes at $t = 100$ in the insets in figure 9. It is seen that the droplet deforms the largest for highest confinement ratio $(a/\bar{H} = 0.4)$ due to the strong effect of the imposed flow. The droplet for this case is found to be stretched in the axial direction with a sharp leading edge and a flat trailing edge. The deformation of the droplet is the minimum for $a/\bar{H} = 0.2875$. All the other parameters, $Ca$, $Ma_T$, $Pe_T$, $\xi$, $\omega$, $\lambda$ as well as the initial position of the droplet are kept constant.

## 5.5. Effect of conductivity ratio, $\xi$

We finally show the effect of the thermal conductivity ratio on the lateral migration of the droplet in figure 10. It can be seen from the figure that with increase in $\xi$, the droplet reaches a steady state position which is closer to the channel centerline in comparison to a system with a lower value of $\xi$. For a higher value of $\xi$, the range of temperature variation or the temperature gradient along the droplet interface decreases, which thus results in a decrease of the thermal Marangoni stress. Thus the imposed flow has a higher dominating effect on the droplet migration as compared to the temperature induced Marangoni stress. As a result, for a system with higher $\xi$, the hydrodynamic force due to the imposed flow drives the droplet more closer to the channel centerline. This reasoning is in direct agreement with the work done by Young et al,(Young et al. 1959) where they studied the thermocapillary migration of a bubble (as well as that of a droplet) in a channel of circular cross section. The expression for the dimensional droplet migration velocity given by Young et al is(Young et al. 1959)

$$\bar{\mathbf{u}}_{YGB} = -\frac{2}{3}\frac{1}{(3+2\lambda)}\left(\frac{3\mu_e \beta a T_c}{2+\xi}\right)\mathbf{e}_y, \qquad (47)$$

where $a$ is the radius of drop and $\mathbf{e}_y$ is the unit vector along positive $y$-direction. In the above expression for the migration velocity of the droplet, we have considered the special case of a neutrally buoyant droplet $(\rho_i = \rho_e)$. From equation (47), it is seen that with increase in $\xi$, the droplet migration velocity, induced solely due to thermocapillary effect, reduces which indicates that the droplet migrates more closer the channel centerline. This theoretical prediction matches our numerical outcome. In figure 10 we show the time variation of the lateral position of the droplet for three different values of conductivity ratio, $\xi = 0.5, 1, 2$. The other parameters namely, $a/\bar{H}$, $Ca$, $Ma_T$ as well as the initial position of the droplet is kept the same for each case. As explained above, for the system with $\xi = 2$, the droplet reaches the steady state position closest to the channel centerline.

The effect of thermal Péclet number on the migration of the droplet was also investigated. However, the droplet migration was seen to be almost unaffected with any change in $Pe_T$. As a result, no further attempt has been made to show the variation of the droplet lateral position with time, for different values of $Pe_T$. In all the simulations performed in this paper we have taken $Pe_T = 20$.

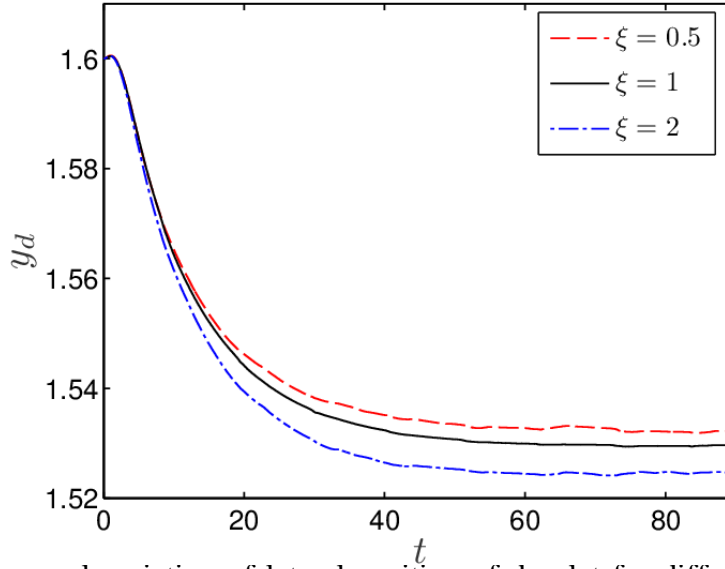

FIGURE 10. Temporal variation of lateral position of droplet for different values of thermal conductivity ratio $\xi = 0.5, 1, 2$. The other parameters are $\omega = \lambda = 1$, $Re = 4$, $Pe_T = 20$, $Ca = 0.3$, $a/\bar{H} = 0.333$ and $Ma_T = 0.1$. The droplet is initially placed above the channel centerline with $y_d(t=0) = 1.6$.

## 6. Conclusion

In the present paper, we have looked into the droplet migration characteristics in a micro channel under the combined effect of an imposed plane Poiseuille flow and a constant transverse temperature gradient. Thus from our present analysis any steady state transverse position of the droplet can be predicted. This transverse position of the droplet can be altered by changing the different parameters like confinement ratio, temperature gradient or thermal conductivity ratio. We have used an asymptotic approach to obtain the solution for axial as well as cross-stream velocity till $O(Ca)$. We have also used the conservative level set methodology to numerically solve the flow field for arbitrary thermal Péclet number. After a proper analysis of the effects of the various physical parameters, a number of important findings were established which are mentioned below.

- A good qualitative match was found between out theoretical and numerical results. The trend of variation lateral position of the droplet with time was found to be similar in either of the cases.

- For a particular value of $Ca$, increase in the thermal Marangoni number $(Ma_T)$ shifts the steady state position of the droplet further away from the channel centerline. The value of $Ma_T$ for which there is absolutely no lateral migration of the droplet is the critical value of thermal Marangoni number. Beyond the critical Marangoni number, the droplet migrates towards the wall of the channel. Below this value, the droplet migrates towards the center of the channel.

- With increase in the value of $Ca$, the critical Marangoni number increases. In other words, for a constant value of $Ma_T$, a droplet, which was initially migrating towards the channel centerline for a higher value of $Ca$, moves towards the wall for a lower value of $Ca$.

- The steady state transverse position of the droplet was found to be independent of the initial position of the droplet.

- The direction of migration of the droplet can be controlled by adjusting the confinement ratio. A higher confinement ratio forces the droplet to migrate towards the channel center whereas a low enough value of $a/\bar{H}$ causes the droplet to move towards the wall. Also, for a system with a higher $a/\bar{H}$ value, a larger droplet deformation is imminent.

- A higher value of $\xi$ causes the droplet to migrate more closer to the channel centerline as compared to a system with a low value of $\xi$.

**Appendix A: expressions of the constant coefficients as present in equation (33)**

The constant coefficients present in the expression of $O(Ca)$ temperature distribution, $T_i^{(Ca)}, T_e^{(Ca)}$, as shown in equation (33) are given below

$$\left.\begin{aligned}
b_{-2,0}^{(Ca)} &= \frac{3(\xi+5)(\xi-1)(16+19\lambda)(H-2y_d)}{10(\xi+2)^2(1+\lambda)H^2}, \\
b_{-4,0}^{(Ca)} &= -\frac{9(16+19\lambda)(\xi-1)(H-2y_d)\xi}{10(\xi+2)(3\xi+4)(1+\lambda)H^2}, \\
b_{-3,1}^{(Ca)} &= -\frac{8(\xi+3)(\xi-1)(10+11\lambda)}{35(\xi+2)(2\xi+3)(1+\lambda)H^2}, \\
b_{-4,2}^{(Ca)} &= \frac{3(16+19\lambda)(\xi-1)(H-2y_d)\xi}{20(\xi+2)(3\xi+4)(1+\lambda)H^2},
\end{aligned}\right\} \quad (A1)$$

$$\left.\begin{aligned}
a_{1,0}^{(Ca)} &= \frac{9(19\lambda+16)(\xi-1)(-2y_d+H)}{10(\xi+2)^2(1+\lambda)H^2}, \\
a_{3,0}^{(Ca)} &= \frac{6(19\lambda+16)(\xi-1)(-2y_d+H)}{5(\xi+2)(3\xi+4)(1+\lambda)H^2}, \\
a_{2,1}^{(Ca)} &= -\frac{12(10+11\lambda)(\xi-1)}{35(\xi+2)(2\xi+3)(1+\lambda)H^2}, \\
a_{3,2}^{(Ca)} &= -\frac{(19\lambda+16)(\xi-1)(-2y_d+H)}{5(\xi+2)(3\xi+4)(1+\lambda)H^2}.
\end{aligned}\right\} \quad (A2)$$

**Appendix B: validation of our numerical setup**

We first match our numerical result with that obtained by Mortazavi and Tryggvason (2000). They have used the finite difference/front-tracking method to track the droplet interface. In figure 11 we have shown the evolution of the transverse position of the droplet with time for $Re = 4$ and $a/\overline{H} = 0.375$. The viscosity ratio and the density ratio between the droplet phase and the carrier phase are taken to be unity. As can be seen from the figure below, there is good match between the our result and the result obtained by Mortazavi and Tryggvason.

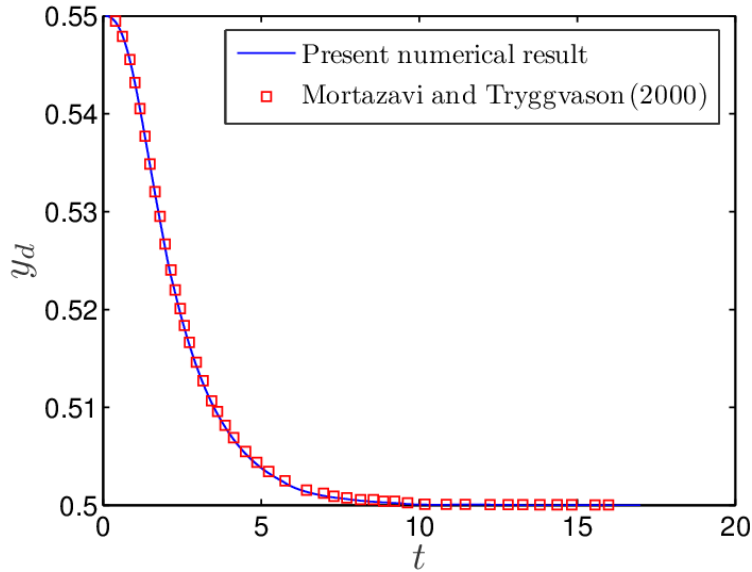

FIGURE 11. Variation of transverse position of the droplet with time. The different parameters involved in the above plot are $Re = 4$, $a/H = 0.375$, $\omega = 1$, $\lambda = 1$.

Next, we match the time evolution of the axial droplet migration velocity with that obtained by Wu and Hu (2011) as well as Nas and Tryggvason (2003) when an axial temperature gradient is applied in the absence of any imposed flow. Both the papers used the front-tracking method to solve for the flow field. However in our present work we have used the conservative level set approach. As can be seen from the from figure 12, there is a better match between our result with that obtained by Nas and Tryggvason as compared to that between the results of Wu and Hu (2011) and Nas and Tryggvason (2003). The different parameters used for this simulation, as per Nas and Tryggvason are $Re = 5$, $Ca = 0.01667$, $Ma_T = 20$, $\omega = \lambda = \xi = 2$. Only for this particular case of validation, we haven't considered the droplet to be neutrally buoyant. For all the other simulations carried out throughout this paper, we have taken $\omega = 1$. The computational domain for the purpose of this validation was taken as $4 \times 8$ with $128 \times 256$ grid points that is 32 grid points per droplet radius. In figure 12 $\mathbf{u}(y_d = 0)$ is the axial droplet velocity when placed at the channel centerline.

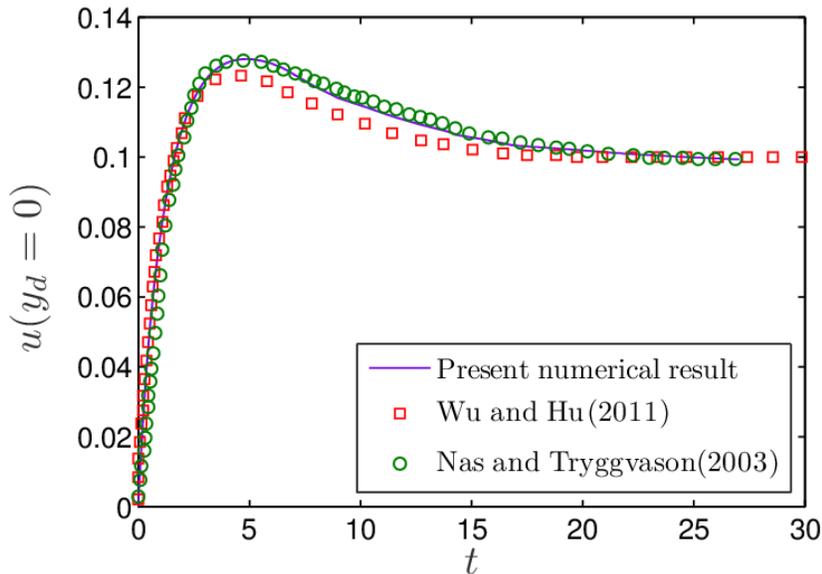

FIGURE 12. Variation of droplet migration velocity with time when the droplet is placed at the centerline of flow with a linearly increasing temperature field in the axial direction. The results obtained by Nas and Tryggvason (2003) and Wu and Hu (2011) is plotted along with the result obtained from the present numerical approach. The different parameters used above are: $Ca = 0.01667$, $Re = 5$, $Ma_T = 20$, $\xi = \lambda = \omega = 2$.

**Appendix C: grid independency check**

We next carry out a grid independency check to ensure that our results remains the same irrespective of the element size chosen. We, in our numerical simulations have chosen a maximum element size of 0.03667. We, thus, further carry out full scale simulations (figure 13) to show the time variation of the lateral position of the droplet for $(Ca, Ma_T) = (0.3, 0.1)$ with maximum element sizes of 0.04 and 0.03. All the other parameters involved in these simulations are kept constant and are given by: $a/\bar{H} = 0.333$, $\text{Re} = 4$ and $Pe_T = 20$. As seen from figure 13, there is a good match between the results for the three different runs corresponding to the maximum element size chosen.

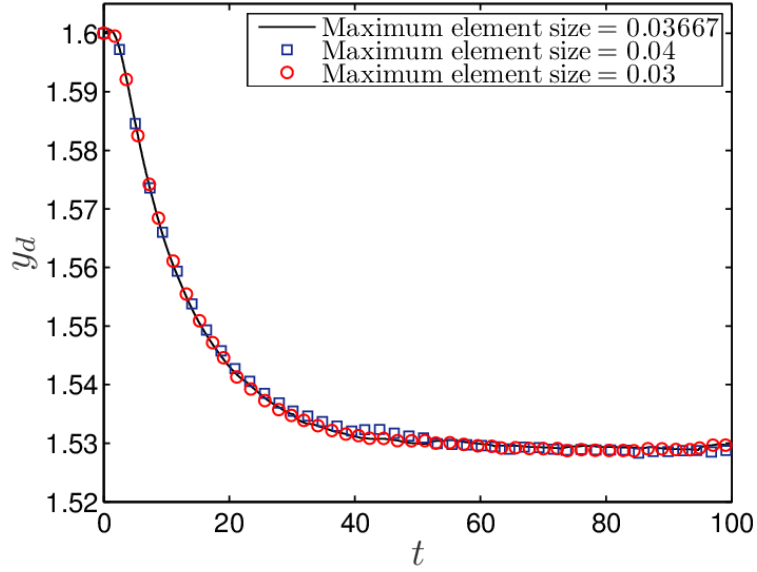

FIGURE 13. Grid independency check done for three different maximum element sizes (0.03, 0.036667, 0.04) corresponding to $Ca = 0.3$, and $Ma_T = 0.1$. The other parameters are $Re = 4$, $Pe_T = 20$, $H = 3$, and $L = 9$.

**Appendix D: expression of the constants in equation (46)**

The expression of the constants $c$ and $e$ present in the equation (46) are given below

$$\varpi = -\frac{16}{105} \frac{\left(198\lambda^5 - 1242\lambda^4 - 7327\lambda^3 - 6292\lambda^2 + 1843\lambda + 2320\right)}{H^4(\lambda+1)^2(2+3\lambda)^2(\lambda+4)} Ca,$$

$$\upsilon = -\left[\begin{array}{c} \dfrac{2Ma_T}{(\xi+2)(3\lambda+2)} \\ +\dfrac{8}{105}\dfrac{\left(198\lambda^5 - 1242\lambda^4 - 7327\lambda^3 - 6292\lambda^2 + 1843\lambda + 2320\right)}{H^4(\lambda+1)^2(2+3\lambda)^2(\lambda+4)}\left\{(H-2e)+2y_d\right\}Ca \end{array}\right] \quad \text{(D1)}$$